\documentclass[aps,prl,reprint,groupedaddress,superscriptaddress]{revtex4-1}
\usepackage{amssymb}
\usepackage{amsmath}
\usepackage{graphicx}
\usepackage{epstopdf}
\usepackage{verbatim}

\begin{document}

\title{Pressure-Induced Modification of Anomalous Hall Effect in Layered Fe$_3$GeTe$_2$}
\author{Xiangqi Wang$^\ddagger$}
    \affiliation{CAS Key Laboratory of Strong-Coupled Quantum Matter Physics and Department of Physics, University of Science and Technology of China, Hefei, Anhui 230026, China}
\author{Zeyu Li$^\ddagger$}
    \affiliation{CAS Key Laboratory of Strong-Coupled Quantum Matter Physics and Department of Physics, University of Science and Technology of China, Hefei, Anhui 230026, China}
    \affiliation{ICQD, Hefei National Laboratory for Physical Sciences at Microscale, and Synergetic Innovation Centre of Quantum Information and Quantum Physics, University of Science and Technology of China, Hefei, Anhui 230026, China}
\author{Min Zhang}
    \affiliation{CAS Key Laboratory of Strong-Coupled Quantum Matter Physics and Department of Physics, University of Science and Technology of China, Hefei, Anhui 230026, China}
\author{Tao Hou}
    \affiliation{CAS Key Laboratory of Strong-Coupled Quantum Matter Physics and Department of Physics, University of Science and Technology of China, Hefei, Anhui 230026, China}
    \affiliation{ICQD, Hefei National Laboratory for Physical Sciences at Microscale, and Synergetic Innovation Centre of Quantum Information and Quantum Physics, University of Science and Technology of China, Hefei, Anhui 230026, China}
\author{Jinggeng Zhao}
    \affiliation{Department of Physics, and Natural Science Research Center, Harbin Institute of Technology, Harbin 150080, China}
\author{Lin Li}
    \affiliation{CAS Key Laboratory of Strong-Coupled Quantum Matter Physics and Department of Physics, University of Science and Technology of China, Hefei, Anhui 230026, China}
\author{Azizur Rahman}
\author{Zilong Xu}
\author{Junbo Gong}
    \affiliation{CAS Key Laboratory of Strong-Coupled Quantum Matter Physics and Department of Physics, University of Science and Technology of China, Hefei, Anhui 230026, China}
\author{Zhenhua Chi}
    \affiliation{Key Laboratory of Materials Physics, Institute of Solid State Physics, Chinese Academy of Sciences, Hefei 230031, China}
\author{Rucheng Dai}
\author{Zhongping Wang}
    \affiliation{The Center of Physical Experiments, University of Science and Technology of China, Hefei 230026, China}
\author{Zhenhua Qiao}
    \email[Correspondence author:~]{qiao@ustc.edu.cn}
    \affiliation{CAS Key Laboratory of Strong-Coupled Quantum Matter Physics and Department of Physics, University of Science and Technology of China, Hefei, Anhui 230026, China}
    \affiliation{ICQD, Hefei National Laboratory for Physical Sciences at Microscale, and Synergetic Innovation Centre of Quantum Information and Quantum Physics, University of Science and Technology of China, Hefei, Anhui 230026, China}
\author{Zengming Zhang}
    \email[Correspondence author:~]{zzm@ustc.edu.cn}
    \affiliation{CAS Key Laboratory of Strong-Coupled Quantum Matter Physics and Department of Physics, University of Science and Technology of China, Hefei, Anhui 230026, China}
    \affiliation{The Center of Physical Experiments, University of Science and Technology of China, Hefei 230026, China}

\begin{abstract}
  We systematically investigate the influence of high pressure on the electronic transport properties of layered ferromagnetic materials, in particular, those of Fe$_3$GeTe$_2$. Its crystal sustains a hexagonal phase under high pressures up to 25.9 GPa, while the Curie temperature decreases monotonously with the increasing pressure. By applying appropriate pressures, the experimentally measured anomalous Hall conductivity, $\sigma_{xy}^A$, can be efficiently controlled. Our theoretical study reveals that this finding can be attributed to the shift of the spin--orbit-coupling-induced splitting bands of Fe atoms. With loading compression, $\sigma_{xy}^A$ reaches its maximal value when the Fermi level lies inside the splitting bands and then attenuates when the splitting bands float above the Fermi level. Further compression leads to a prominent suppression of the magnetic moment, which is another physical cause of the decrease in $\sigma_{xy}^A$ at high pressure. These results indicate that the application of pressure is an effective approach in controlling the anomalous Hall conductivity of layered magnetic materials, which elucidates the physical mechanism of the large intrinsic anomalous Hall effect.
\end{abstract}

\maketitle

\textit{Introduction---.} Anomalous Hall effect has been one of the most attractive but unsolved topics within the condensed matter community ever since its experimental discovery~\cite{nagaosa2010anomalous}. In general, anomalous Hall effect is closely related to the material magnetization, and its fundamental origin in different materials is debatable: it is commonly rationalized as being extrinsic disorder-induced effects (e.g., skew-scattering and side jump) or intrinsic Berry-phase effect~\cite{chen2014anomalous,nakatsuji2015large,ruan2016symmetry,nayak2016large,wang2018large,kim2018large}. Nevertheless, new types of anomalous Hall effects in material systems besides those in ferromagnetic materials (i.e., topological Hall effect in non-collinear antiferromagnetic materials~\cite{chen2014anomalous,nakatsuji2015large,ruan2016symmetry,nayak2016large} and giant anomalous Hall effect in magnetic semimetals~\cite{wang2018large,kim2018large}) continually update our understanding of such a striking but unclear electronic transport phenomenon. Noteworthily, in these materials, the anomalous Hall effect is not simply related to the magnetization that breaks time-reversal symmetry; thus, the effect cannot be understood from conventional formation mechanisms. An increasing number of studies have demonstrated that the anomalous Hall effect is intimately connected with the intrinsic Berry-phase effect from spin--orbit couplings. Recent works~\cite{onoda2008quantum,nagaosa2010anomalous} illustrate that  anomalous Hall conductivity $\sigma_{ xy}^{\rm A}$ is dominated by $\sigma_{xy}^{\rm skew}$ from the skew-scattering in the high-conductive regime of $\sigma_{xx}>10^6 (\Omega\cdot cm)^{-1}$. When $\sigma_{xx}$ drops to $10^4\sim10^6 (\Omega\cdot cm)^{-1}$, the dominant part is replaced by intrinsic $\sigma_{ xy}^{\rm int}$.

Fe$_3$GeTe$_2$ is an itinerant van der Waals ferromagnet with a relatively high Curie temperature, $T_c$ $\sim$ 150--220 K, depending on the Fe occupancy level ~\cite{yi2016competing,may2016magnetic,liu2017wafer,Liu2018}. It is crystallized as a hexagonal structure with space group \emph{P}$_{63}$/\emph{mmc}. The layered structure along the \emph{c} direction contains Fe$_3$Ge slabs sandwiched by Te layers, which are coupled by weak van der Waals interaction \cite{chen2013magnetic,deiseroth2006fe3gete2}. This has attracted consistent attention from researchers due to its unusual properties, e.g., strong out-of-plane anisotropy, Kondo lattice behavior, and the electron correlation effect ~\cite{may2016magnetic,zhang2018emergence,zhu2016electronic}. Recently, it has been reported that two-dimensional ferromagnetism can still be sustained at high T$_c$ for the thin flakes down to the monolayer limit, where the Hall effect measurement plays a key role in characterizing the two-dimensional ferromagnetism~\cite{deng2018gate,fei2018two,liu2017wafer}. Furthermore, Fe$_3$GeTe$_2$ was considered as a ferromagnetic topological nodal semimetal candidate, where the nodal line is tunable via spin orientation owing to spin--orbit coupling, and it produces a large Berry curvature to generate a relatively large anomalous Hall current~\cite{kim2018large}. Together with other magnetic topological semimetals (e.g., Co$_3$Sn$_2$S$_2$ \cite{wang2018large} and GdPtBi \cite{shekhar2018anomalous}), this provides an ideal platform for exploring the anomalous Hall effect as well as the quantum anomalous Hall effect in a two-dimensional limit.

\begin{figure}[htb]
  \includegraphics[width=8.6cm]{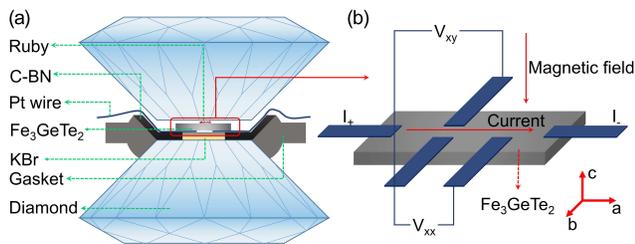}%
  \caption{(a) Schematic of a high-pressure diamond anvil cell electrical-properties measurement system. Cubic-BN was used to insulate Pt wires from a gasket. KBr was used as a solid pressure-transmitting medium. The pressure was identified via the ruby fluorescence shift. (b) Schematic of a standard Hall bar wiring method used inside a diamond anvil cell.}
  \label{FIG1}
\end{figure}

High pressure, as an externally manipulating parameter, is a viable tool for controlling the crystal lattice and the corresponding electronic states~\cite{chi2018superconductivity,lin2018pressure}. To our knowledge, although there have been various methods utilized to manipulate the anomalous Hall effect, no experimental observation of high-pressure-modulated anomalous Hall effect has been reported so far. In this Letter, we demonstrate that the crystal structure, band structure, and the electronic transport properties of layered Fe$_3$GeTe$_2$ can be substantially affected by applying pressure. In particular, we find that the anomalous Hall effect can dramatically change by increasing the pressure within our measured range, which has been shown to originate from the relative shift of spin--orbit-splitting bands from the Fermi level using density functional theory. This work indicates that high pressure provides an effective external approach to reliably control the anomalous Hall effect in ferromagnetic materials, and it offers a new route for exploring the fundamental mechanisms of the anomalous Hall effect.

\begin{figure}[htbp]
\includegraphics[width=8.6cm]{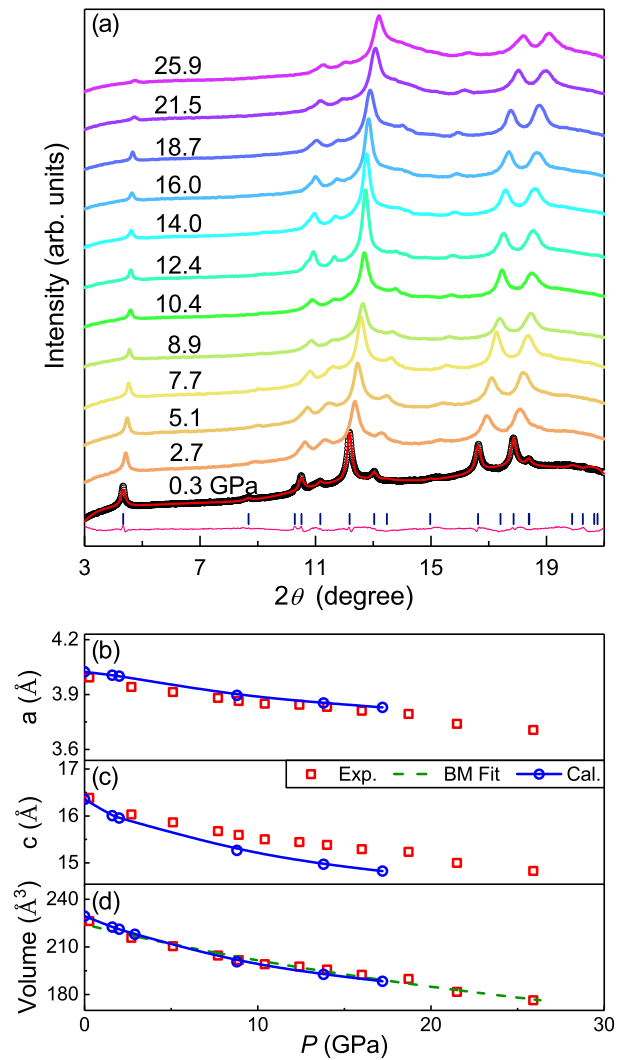}%
\caption{(a) High-pressure synchrotron X-ray diffraction (XRD) patterns for Fe$_3$GeTe$_2$ from 0.3 GPa to 25.9 GPa at room temperature. The Rietveld refinement of diffraction pattern at 0.3 GPa is shown as a representative. (b--d) exhibit experimental and calculated lattice constants \emph{a}, \emph{c}, and unit-cell volume as a function of pressure, respectively. The green line shows the Birch--Murnaghan state equation fitting of refined P-V data from experimental results.}
\label{FIG2}
\end{figure}

\textit{Methods---.} Single crystals of Fe$_3$GeTe$_2$ were grown via chemical vapor transport method~\cite{chen2013magnetic}. X-ray diffraction (XRD) and energy dispersive spectroscopy experiments were performed to confirm the high structural and compositional quality. High-pressure synchrotron XRD measurement was performed at beamline 15U1 of the Shanghai Synchrotron Radiation Facility (SSRF). A non-magnetic Cu-Be alloy screw-pressure-type diamond anvil cell was employed to perform a high-pressure transport experiment on a magnet-cryostat Physical Property Measurement System. The schematic plot of diamond anvil cell electrical-transport measurement device is displayed in Fig.~\ref{FIG1}. The density-functional-theory-based \emph{ab initio} calculation methods were used to reveal the physical origin of the varying anomalous Hall effect under high pressure. Details of sample growth, characterization, high-pressure transport and synchrotron XRD experiments are included in the Supplementary Materials.

\begin{figure}[htb]
  \includegraphics[width=8.6cm]{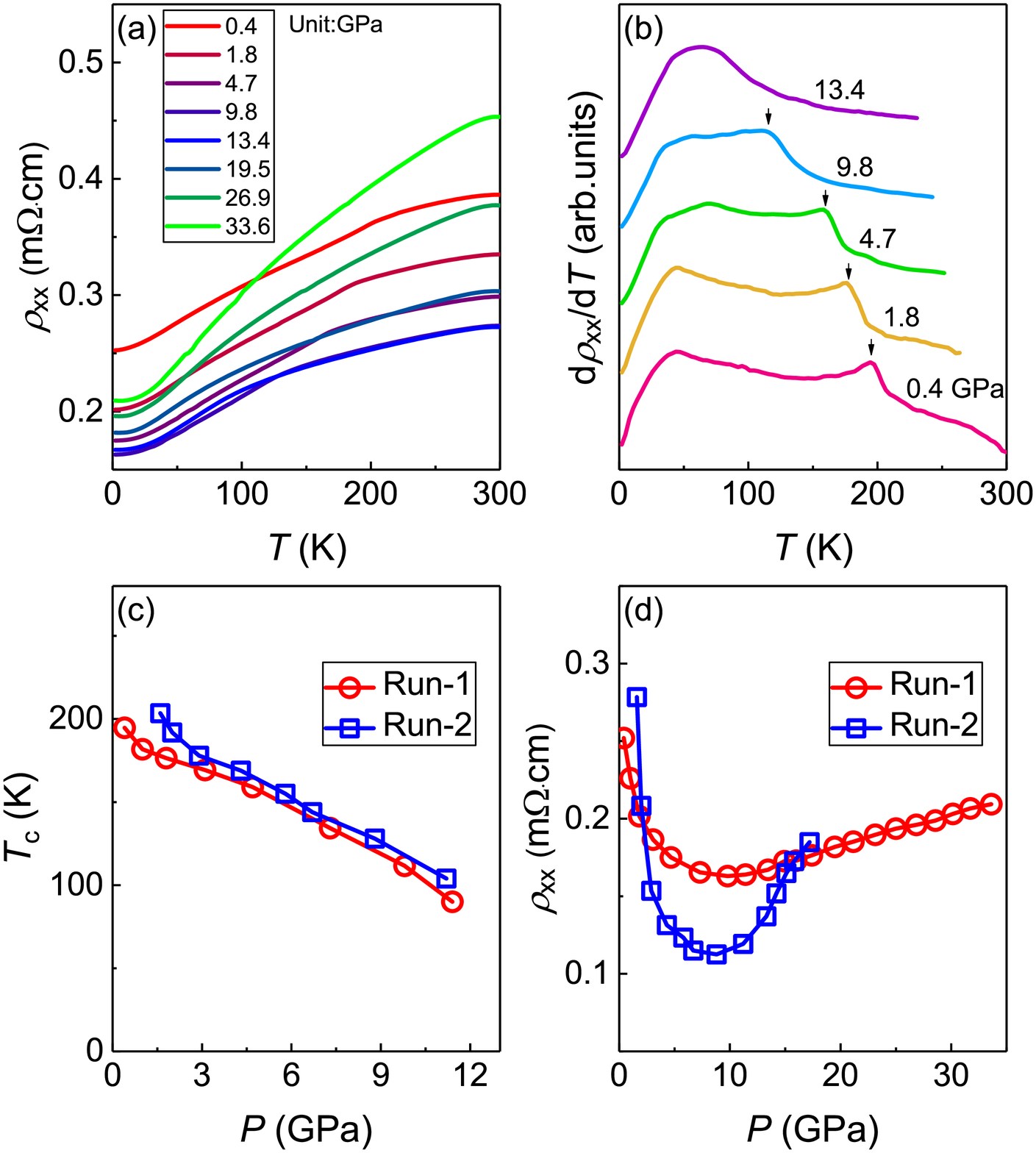}%
  \caption{Electronic transport properties of Fe$_3$GeTe$_2$ under high pressure. (a) Temperature-dependent longitudinal resistivity $\rho_{xx}$ from 0.4 GPa to 33.6 GPa in Run-1. (b) The corresponding temperature derivative of  $\rho_{xx}$ below 13.4 GPa. The curves have been subsequently offset for a clearer view. The arrows point toward the corresponding positions of ferromagnetic transition. Only partial results are shown in the figure for a clear view. Full results of Run-1 and Run-2 can be found in the Supplementary Materials Fig. S (6, 7). (c) The derived Curie temperature as a function of pressure in both Run-1 and Run-2. (d) Pressure-dependent $\rho_{xx}$ at 2 K in both Run-1 and Run-2.}
  \label{FIG3}
\end{figure}

\textit{Structural Stability---.} The structural stability of Fe$_3$GeTe$_2$ was studied by using the \emph{in situ} high-pressure synchrotron angle-dispersive XRD technique. Figure~\ref{FIG2}(a) displays the XRD patterns of Fe$_3$GeTe$_2$ under pressure up to 25.9 GPa at room temperature. All diffraction peaks are well indexed with hexagonal phase structure, and thye gradually shift to a larger angle with increasing pressure, indicating the shrinkage of the lattice upon compression. Surprisingly, no new peak arises, indicating the structural stability of Fe$_3$GeTe$_2$ within the pressure range. A refinement result of the XRD pattern at 0.3 GPa is also displayed as a comparison. The detailed Rietveld refinement results are presented in Table S1.  Figures~\ref{FIG2}(b)--(d) show the refined lattice constants \emph{a}, \emph{c}, and unit-cell volume, together with the \emph{ab initio} calculated results. The refined unit-cell volume against pressure is well-fitted by the Birch--Murnaghan state function, as displayed in Fig.~\ref{FIG2}(d). The layer arrangement leads to the facile compression along the \emph{c} axis, as seen in Fig.~\ref{FIG2}(c). Interestingly, the calculated \emph{c} at different pressures are notably smaller than the experimental values. This reflects that the numerical calculation over-estimates the pressure-induced lattice contraction, which may originate from the limit of computational description of the van der Waals interaction~\cite{vilaplana2011structural}.

\textit{Electronic Transport---.} To investigate the influence of high pressure on electronic transport properties, two runs of resistivity measurement were performed in Fe$_3$GeTe$_2$ single crystals. Figure~\ref{FIG3}(a) displays the longitudinal resistivity $\rho_{xx}$ as a function of temperature at selected pressures in Run-1. The full results of Run-1 and Run-2 can be found in Figs.~S6 and S7. One can observe that this is metallic under all applied pressures, and a remarkable kink representing a phase transition from ferromagnetic to paramagnetic~\cite{chen2013magnetic,deiseroth2006fe3gete2} can be found at pressures below 13.4 GPa within the temperature range from 100 to 200 K. The Curie temperature $T_{\rm C}$ can be derived from the temperature derivative of $\rho_{xx}$, as displayed in Fig.~\ref{FIG3}(b). The remarkable peak around 195 K corresponds to $T_{\rm C}$ at 0.4 GPa, which gradually decreases with increasing pressure before finally becoming indistinguishable for pressures higher than 13.4 GPa. Figure~\ref{FIG3}(c) plots $T_{\rm C}$ as a function of pressure at both Run-1 and Run-2. It shows that $T_{\rm C}$ decreases linearly with a decaying rate of 9.2~K/GPa. This indicates the suppression of ferromagnetic order due to the lattice-shrinkage-induced variance of spin--lattice coupling~\cite{MnP,CGT}. Figure~\ref{FIG3}(d) plots $\rho_{xx}$ as a function of pressure at fixed 2 K in both Run-1 and Run-2, where $\rho_{xx}$ exhibits a clear upturn for pressures over 8 $\sim$ 9 GPa. A similar observation was reported in other topological systems~\cite{zhou2016pressure,zhou2016pressure1,tafti2017tuning,TITE2}. This arises from the suppression of charge-carrier mobility, which may attributed to the pressure-induced inhomogenous contraction of the lattice and non-hydrostatic condition due to the employment of the solid pressure-transmitting medium KBr.

\begin{figure}[tbp]
  \includegraphics[width=8.5cm]{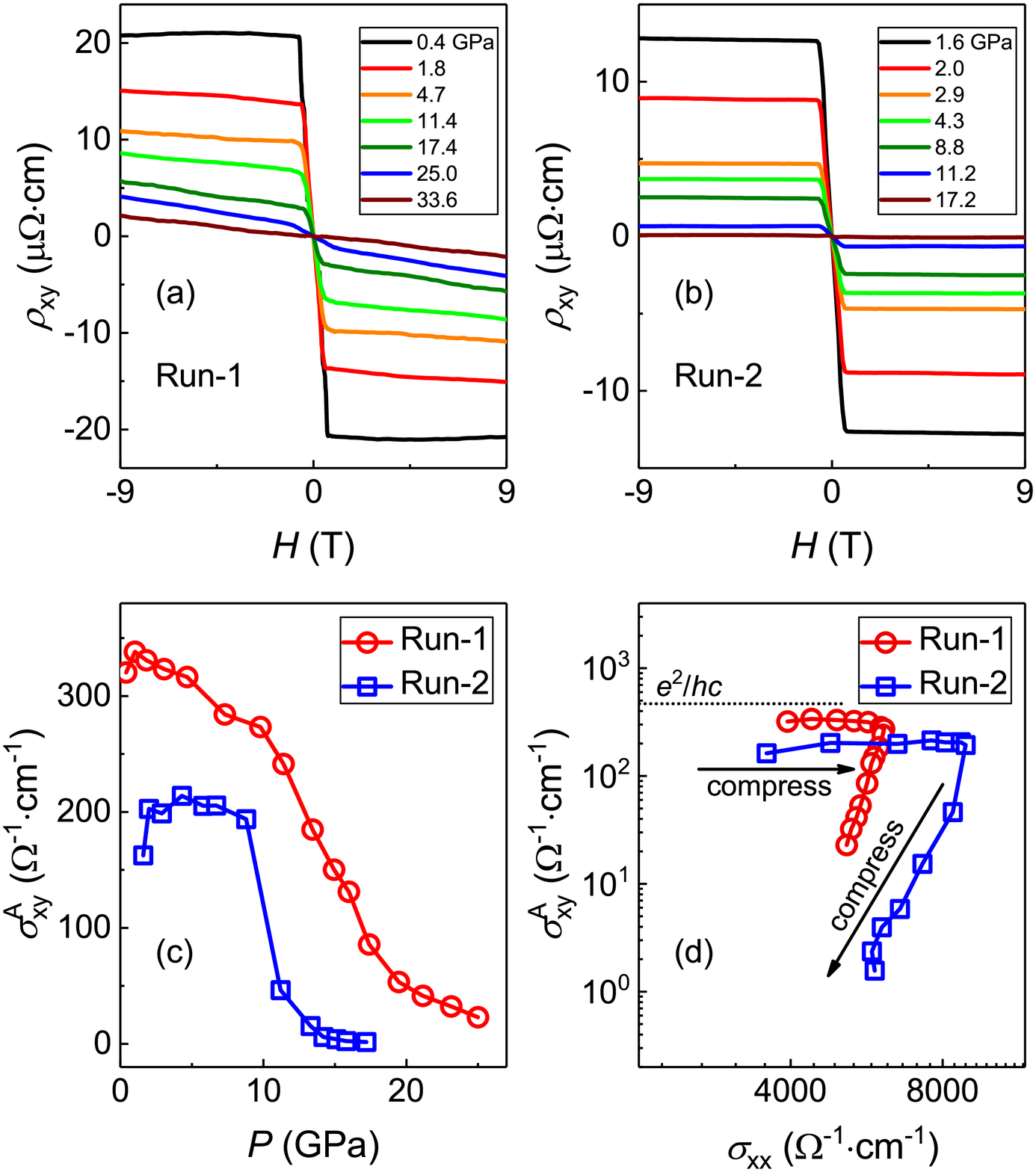}%
  \caption{High-pressure-induced tunable anomalous Hall effect in Fe$_3$GeTe$_2$. (a, b) Field dependent $\rho_{xy}$ at selected pressures in Run-1 and Run-2, respectively. The temperature is kept at 2 K. Full results are provided in Supplementary Materials Fig. S8. (c) Pressure-dependent anomalous Hall conductivity $\sigma_{xy}^A$  in Run-1 and Run-2. (d) $\sigma_{xy}^A(P)$ as a function of $\sigma_{xx}(P)$ for two runs. The arrows indicate the variation of $\sigma_{xy}^A$ against $\sigma_{xx}$ with increasing pressure. The dotted line shows the expected value of $\sigma_{xy}^A$ for quantum anomalous Hall effect. }
\label{FIG4}
\end{figure}

Figures~\ref{FIG4}(a) and \ref{FIG4}(b) display the measured Hall resistivity $\rho_{xy}$ as a function of a magnetic field at different pressures. $\rho_{xy}$ increases dramatically as the magnetic field increases and saturates at $\sim$0.5 T, exhibiting typical features of anomalous Hall effect~\cite{nayak2016large,wang2018large,kim2018large}. In addition, the saturation value of $\rho_{xy}$ decreases monotonically when the pressure increases, while the saturated magnetic field remains almost unchanged. The anomalous Hall conductivity can be obtained from
\begin{equation}
\sigma_{xy}=\rho_{xy}/({\rho^2_{xx}}+{\rho^2_{xy}}).
\end{equation}
Figure~\ref{FIG4}(c) displays the corresponding Hall conductivity as a function of pressure. One can see that $\sigma^{A}_{xy}$ shows a tiny increase under pressure below 2 GPa and a slight decrease with pressures below 8 GPa. With further increasing pressure, the Hall conductivity exhibits a significant drop, until it vanishes. Figure~\ref{FIG4}(d) displays the relation between $\sigma_{xy}^A(P)$ and $\sigma_{xx}(P)$. $\sigma_{xx}$ is in the order of $10^3 (\Omega\cdot cm)^{-1}$, close to the moderately dirty regime, indicating that the intrinsic mechanism dominates the contribution of anomalous Hall effect~\cite{nagaosa2010anomalous,onoda2008quantum}. Below 8$\sim$9 GPa, $\sigma_{xy}^A$ is generally constant against $\sigma_{xx}$ with increasing pressure, consistent with the typical expectation of intrinsic mechanism~\cite{onoda2008quantum,Onoda2006intrinsic,Miyasato2007crossover}. $\sigma_{xy}^A$ in this regime is comparable with the theoretical value of the quantum anomalous Hall effect $e^2/hc$, where \emph{c} represents the interlayer distance~\cite{kim2018large}. With further compression, $\sigma_{xy}^A$ decreases rapidly.

To elucidate the relation between Hall conductivity and external pressure, we explored the band structure of Fe$_3$GeTe$_2$ while being subjected to external pressure by using the first principles calculation methods (e.g., Vienna \emph{ab initio} Simulation Package~\cite{kresse1996efficient}). The projector augmented wave method~\cite{blochl1994projector} and generalized gradient approximation of Perdew--Burke--Ernzerhof parameters~\cite{perdew1996generalized} were employed to describe the system. The Hall conductivity was obtained by the interpolation of maximally localized Wannier functions, constructed in the Wannier90 package~\cite{mostofi2008wannier90,wang2006ab}. Details of computational methods can be found in the Supplementary Materials.

Figures~\ref{FIG5}(a)-\ref{FIG5}(d) display the evolution of band structures of Fe$_3$GeTe$_2$ with both spin--orbit coupling and ferromagnetism for different pressures, i.e., 0, 0.9, 6.9, and 11.2 GPa. Colors are used to highlight the Fe-dominated bands near the K point (like the general Rashba model with ferromagnetism), which mainly contribute to the Hall conductivity. It is well-known that the avoided band crossing at K point originates from the joint interaction between spin--orbit coupling and ferromagnetism, and maximum Hall conductivity is achieved when the Fermi level lies within the gap between the two colored bands at K point. Hall conductivity can be obtained by integrating the Berry curvature below the Fermi for all occupied bands, and the Berry curvature can be calculated as follows:
\begin{equation}
\Omega_{n,\alpha\beta}(k)=-2{\rm Im}\sum_{m\not=n}\frac{v_{nm,\alpha}(k)v_{mn,\beta}(k)}{[\omega_m(k)-\omega_n(k)]^2},
\end{equation}
where $\omega_n(k)=\varepsilon_{nk}/\hbar$ and $v_{nm,\alpha}(k)=\left \langle \Psi_{nk}| \hat{v}_{\alpha}|\Psi_{mk}\right \rangle$ is the matrix elements of velocity operators~\cite{wang2006ab}.

\begin{figure}[tbp]
  \includegraphics[width=8.6cm]{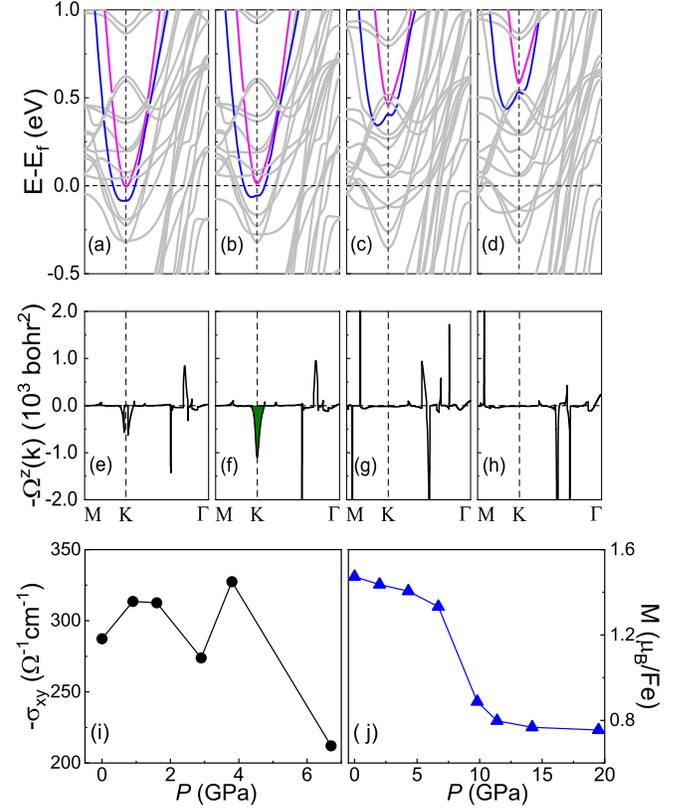}
  \caption{(a--d) Electronic band structure along the symmetry lines at various pressures: (a) 0 GPa, (b) 0.9 GPa, (c) 6.9 GPa, and (d) 11.2 GPa. The colored bands indicate the spin--orbital coupling splitting bands. As expected, they float away from the Fermi level with increasing pressure. (e--h) Corresponding Berry curvature along the symmetry lines. The shaded area indicates the large contribution near the K point. (i) The calculated intrinsic anomalous Hall conductivity as a function of pressure at 0 K. (j) The calculated magnetic moment of an Fe atom as a function of pressure at 0 K.}
\label{FIG5}
\end{figure}

In the absence of pressure, $E_{\rm F}$ is a slightly higher than the minimum of upper splitting band. When the pressure increases, the splitting bands gradually shift up and eventually enter the unoccupied region. Thus, $E_{\rm F}$ will sweep across the whole gap. At the beginning, the upper and lower bands contribute to opposite Berry curvatures, resulting in a misshapen curvature around K point [See Fig.~\ref{FIG5}(e)]. With increasing pressure, $E_{\rm F}$ moves into the splitting gap, with a sizable Berry curvature emerging [as highlighted in Fig.\ref{FIG5}(f)]. As a consequence, the resulting $\sigma_{xy}$ first gradually increases and subsequently maintains a relatively high value. With the continuous increase of pressure, the two splitting bands float away from $E_{\rm F}$, creating a vanishing Berry curvature and Hall conductivity [See Fig.~\ref{FIG5}(g) and \ref{FIG5}(h)]. As only the bands below $E_{\rm F}$ contribute to the Hall conductivity, a conspicuous monotonous dropping of Hall conductivity naturally emerges. Therefore, the evolution of band structure under pressures provides a clear physical understanding of the response of anomalous Hall conductivity to pressure.

For further confirmation, we calculated the Hall conductivity at different pressures [See Fig.~\ref{FIG5}(i)]. At 0 GPa, $\sigma_{xy}$=-287 $(\Omega\cdot cm)^{-1}$, which agrees well with the experimental results and previous reports~\cite{kim2018large,wang2017anisotropic}. $\sigma_{xy}$ increases and maintains a relatively large value when $E_{\rm F}$ lies in the splitting band gap; it finally decreases when the splitting bands float above $E_{\rm F}$. All the numerical findings are qualitatively consistent with the experimental result. Furthermore, it is known that local density approximation can be used to evaluate the magnetic moment within the range of experimental values~\cite{Zhuang2016lda}. Our calculations show that as the pressure increases, prominent suppression of the magnetic moment occurs [See Fig.~\ref{FIG5}(j)], i.e., along with the increase in pressure, the magnetic moment of Fe decreases slowly and then drops quickly after 6.7 GPa. Because in ferromagnets $\sigma_{xy}$ is proportional to the magnetic moment (e.g., bcc Fe and Co$_3$Sn$_2$S$_2$~\cite{wang2018large,Suzuki2017Fe}), one can consider the sudden drop in the magnetic moment to be the other physical origin of the decrease in Hall conductivity at high pressures.

\textit{Summary---.} We studied the electronic properties of the layered ferromagnetic Fe$_3$GeTe$_2$ at high pressures. In the absence of external pressure, its anomalous Hall conductivity is relatively large owing to the contribution from the Rashba-like bands of Fe. We show that the large anomalous Hall effect in Fe$_3$GeTe$_2$ is efficiently controllable by applying external pressure. For low pressures, the decrease in Hall conductivity can be attributed to the  Fe-dominated Rashba-like bands floating up; whereas for high pressure, the other physical origin is the prominent suppression of the magnetic moment of Fe. Our findings demonstrate that high-pressure can play an efficient role in tuning the anomalous Hall effect in magnetic materials and offer a new route for further exploration of the physical mechanism of the anomalous Hall effect.

\textit{Acknowledgments---.} This work was financially supported by the Science Challenge Project (TZ2016001), the National Natural Science Foundation of China (11672273, 11474265, and 11704366), the National Key Research and Development Program (2016YFA0301700). We are grateful to the beamline 15U1 of the SSRF. We also thank the supercomputing service of AM-HPC and the Supercomputing Center of USTC for providing the high-performance computing resources.

$^\ddagger$ X.Q. Wang and Z.Y. Li contributed equally to this work.

\end{document}